On the inapplicability of a negative-phase-velocity condition as a negative-refraction condition for active materials

**Akhlesh Lakhtakia**
*NanoMM–Nanoenginered Metamaterials Group*
*Department of Engineering Science and Mechanics*
*Pennsylvania State University*
*University Park, PA 16802, USA*

**Tom G. Mackay**
*School of Mathematics*
*University of Edinburgh*
*Edinburgh EH9 3JZ, United Kingdom*

**Joseph B. Geddes III**
*Beckman Institute for Advanced Science and Technology*
*University of Illinois at Urbana-Champaign*
*Urbana, IL 61801, USA*

**Abstract.** A negative-phase-velocity condition derived by Depine and Lakhtakia [*Microwave Opt Technol Lett* 41 (2004) 315] for isotropic, homogeneous, passive, dielectric-magnetic materials is inapplicable as a negative-refraction condition for active materials.

Let us consider an isotropic, homogeneous, dielectric-magnetic material with relative permittivity $\varepsilon = \varepsilon_r + i\varepsilon_i$ and relative permeability $\mu = \mu_r + i\mu_i$; its refractive index $n = \sqrt{\varepsilon\,\mu}$.

Suppose that this material is passive, i.e., $\varepsilon_i > 0$ and $\mu_i > 0$. Plane waves in this material can propagate such that the phase velocity is either co-parallel or anti-parallel with respect to the time-averaged Poynting vector. In the former case, the material is classified as a positive-phase-velocity (PPV) material; in the latter case, as a negative-phase-velocity (NPV) material. A PPV material refracts positively because its refractive index is such that $\text{Re}(n) > 0$, whereas a NPV material refracts negatively because $\text{Re}(n) < 0$ for it [1].

Depine and Lakhtakia [2] showed that a passive material is of the NPV type if and only if

$$\varepsilon_r |\mu| + \mu_r |\varepsilon| < 0. \tag{1}$$

This condition thus identifies if a passive material will refract negatively or not. Is this condition valid also as a negative-refraction (NR) condition if the material is not passive; i.e., $\varepsilon_i < 0$ and/or $\mu_i < 0$?

In order to answer this question, let us choose a purely dielectric material. Then, $\mu = 1$ and the Depine-Lakhtakia condition (1) simplifies to

$$\varepsilon_r + |\varepsilon| < 0 \ . \qquad (2)$$

Regardless of the sign of $\varepsilon_r$, this condition can never be satisfied by any material, passive or active. Therefore, if the Depine-Lakhtakia condition (1) were to be valid as an NR condition for active materials too, then an active, nonmagnetic, dielectric material cannot be of the NR type.

Recent literature contains detailed examination of an active, nonmagnetic, dielectric material with relative permittivity [3,4]

$$\varepsilon(\omega) = 1 + \frac{\alpha_1 \omega_1^2}{\omega_1^2 - (\omega + i\beta_1\omega_1)^2} + \frac{\alpha_2 \omega_2^2}{\omega_2^2 - (\omega + i\beta_2\omega_2)^2}, \qquad (3)$$

where $\alpha_1 = 2.4401$, $\alpha_2 = -0.14348$, $\beta_1 = 0.028571$, $\beta_2 = 0.02$, $\omega_1 = 2.6371 \times 10^{15}$ rad/s, $\omega_2 = 3.7673 \times 10^{15}$ rad/s, and $\omega$ is the angular frequency. With $\lambda_0$ denoting the free-space wavelength, this material is active for $\lambda_0 \in [445, 535]$ nm. The proposers of this material asserted that it is of the NR type in that spectral regime [3]. Also, Skaar argued that one cannot unambiguously determine the refractive index of an active material from its relative permittivity and permeability at one frequency [4]. Independent confirmation of the NR assertion––for example, by the solution of a time–domain problem––was lacking. The time-domain solution of an initial-boundary-value problem confirmed that assertion, as also did the frequency-domain solution of a boundary-value problem wherein $\varepsilon$ appears but not its square root [5]. Since (3) satisfies the condition $\varepsilon_r + |\varepsilon| \geq 0$, it cannot satisfy (2). Hence, the Depine-Lakhtakia condition (1) cannot be used to determine whether or not an *active* material is of the NR type. Let us, however, note that such a determination may not be crucial for many boundary-value problems of practical significance [6].